\documentclass[%
reprint,
superscriptaddress,
showpacs,
 amsmath,amssymb,
 prl
]{revtex4-1}

\usepackage{graphicx}
\usepackage{dcolumn}
\usepackage{bm}
\usepackage{xcolor}

\begin{document}

\title{Adiabatic evolution on a spatial-photonic Ising machine}

\author{D. Pierangeli}
\email{davide.pierangeli@roma1.infn.it}
\affiliation{Dipartimento di Fisica, Universit\`{a} di Roma  ``La Sapienza'', 00185 Rome, Italy}
\affiliation{Institute for Complex System, National Research Council (ISC-CNR), 00185 Rome, Italy}

\author{G. Marcucci}
\affiliation{Dipartimento di Fisica, Universit\`{a} di Roma  ``La Sapienza'', 00185 Rome, Italy}
\affiliation{Institute for Complex System, National Research Council (ISC-CNR), 00185 Rome, Italy}

\author{C. Conti}
\affiliation{Institute for Complex System, National Research Council (ISC-CNR), 00185 Rome, Italy}
\affiliation{Dipartimento di Fisica, Universit\`{a} di Roma  ``La Sapienza'', 00185 Rome, Italy}

\begin{abstract}
Combinatorial optimization problems are crucial for widespread applications but remain difficult to solve on a large scale with conventional hardware.
Novel optical platforms, known as coherent or photonic Ising machines, are attracting considerable attention as accelerators on optimization tasks 
formulable as Ising models.  
Annealing is a well-known technique based on adiabatic evolution for finding optimal solutions in classical and quantum systems made by atoms, electrons,
or photons. Although various Ising machines employ annealing in some form, 
adiabatic computing on optical settings has been only partially investigated.
Here, we realize the adiabatic evolution of frustrated Ising models with $100$ spins programmed by spatial light modulation. We use holographic and optical control to change the spin couplings adiabatically, and exploit experimental noise to explore the energy landscape. Annealing enhances the convergence to the Ising ground state and allows to find the problem solution with probability close to unity. 
Our results demonstrate a photonic scheme for combinatorial optimization in analogy with adiabatic quantum algorithms and enforced by
optical vector-matrix multiplications and scalable photonic technology.
\end{abstract}

\maketitle

\section{Introduction}

Ising machines are physical devices aimed to accelerate the minimization of Ising Hamiltonians. Scalable implementations of Ising machines are of paramount importance because many of the most challenging combinatorial optimization problems in science, engineering, and social life can be cast in terms of an Ising model \cite{Barahona1982, Lucas2014}.
Finding the ground state of the Ising spin system gives the solution to the optimization, but requires resources growing exponentially with the problem size. For this reason, intense research focuses on unconventional architectures that use computational units such as light pulses \cite{Marandi2014}, superconducting \cite{Johnson2011} and magnetic junctions \cite{Datta2019}, electromechanical modes \cite{Mahboob2016},
lasers and nonlinear waves \cite{Brunner2013, Engheta2019, Ghofraniha2015, Pierangeli2017, Khajavikhan2020, Davidson2019_2}, or polariton and photon condensates \cite{Berloff2017, Dung2017}.

Adiabatic computing is a valuable technique to solve combinatorial optimizations by slowly evolving an easy-to-prepare initial configuration towards the ground state of a target Hamiltonian, which encodes the combinatorial problem \cite{Farhi2001, Santoro2006}. Examples are adiabatic quantum computing using nuclear magnetic resonance \cite{Steffen2003} and superconducting gates \cite{Barends2016}, as well as quantum annealing  with superconducting circuits \cite{Boixo2014}, and simulated annealing on CMOS networks \cite{Yamaoka2016}. 
Annealing is a form of adiabatic computing at non-zero temperatures in which classical, quantum, or nonlinear perturbations enable the exploration of the complex energy landscape \cite{Kirkpatrick1983, Das2008, Blais2017, Goto2016}. Quantum annealers designed to solve classical Ising problems succeed in optimization tasks ranging from protein folding \cite{Aspuru-Guzik2012} to prime factorization \cite{Kais2018}.
If there are advantages from entanglement and quantum tunneling is still debated \cite{Lanting2014, Denchev2016, Mandra2017}.
Therefore, recently large interest is also centred on the the realization of non-electronic annealing devices that can exploits classical nonlinear and photonic properties.

Optical Ising machines use multiple frequency or spatial channels to process data at high speed and in parallel. Coherent Ising machines (CIM) employ optical parametric oscillators \cite{McMahon2016, Inagaki2016, Inagaki2016_2}, fiber lasers \cite{Babaeian2019}, or opto-electronic oscillators \cite{Vandersande2019} to solve optimization problems with remarkable performance for hundreds of spins~\cite{Hamerly2019_2}.
These machines exploit a gain-dissipative principle \cite{Kalinin2018, Bello2019, Davidson2019, Bohm2018, Goto2019, Lvovsky2019}: 
the search for the minimum energy configuration is conducted in an upward direction by gradually raising the gain.
Other platforms based on integrated nanophotonic circuits \cite{Roques-Carmes2020, Prabhu2020, Shen2017, Wu2014, Soci2018, Gaeta2020} operate as optical recurrent neural networks that converge to Ising ground states.
Recurrent feedback also has a key role in the recently-demonstrated large-scale photonic Ising machine by spatial light modulation \cite{Pierangeli2019, Pierangeli2020, Kumar2020}. In all these optical settings, 
the possibility of performing a evolution of the machine's parameters to improve the computation remains largely unexplored.

\begin{figure*}[t!]
\centering
\vspace*{-0.2cm}
\hspace*{-0.2cm} 
\includegraphics[width=1.95\columnwidth]{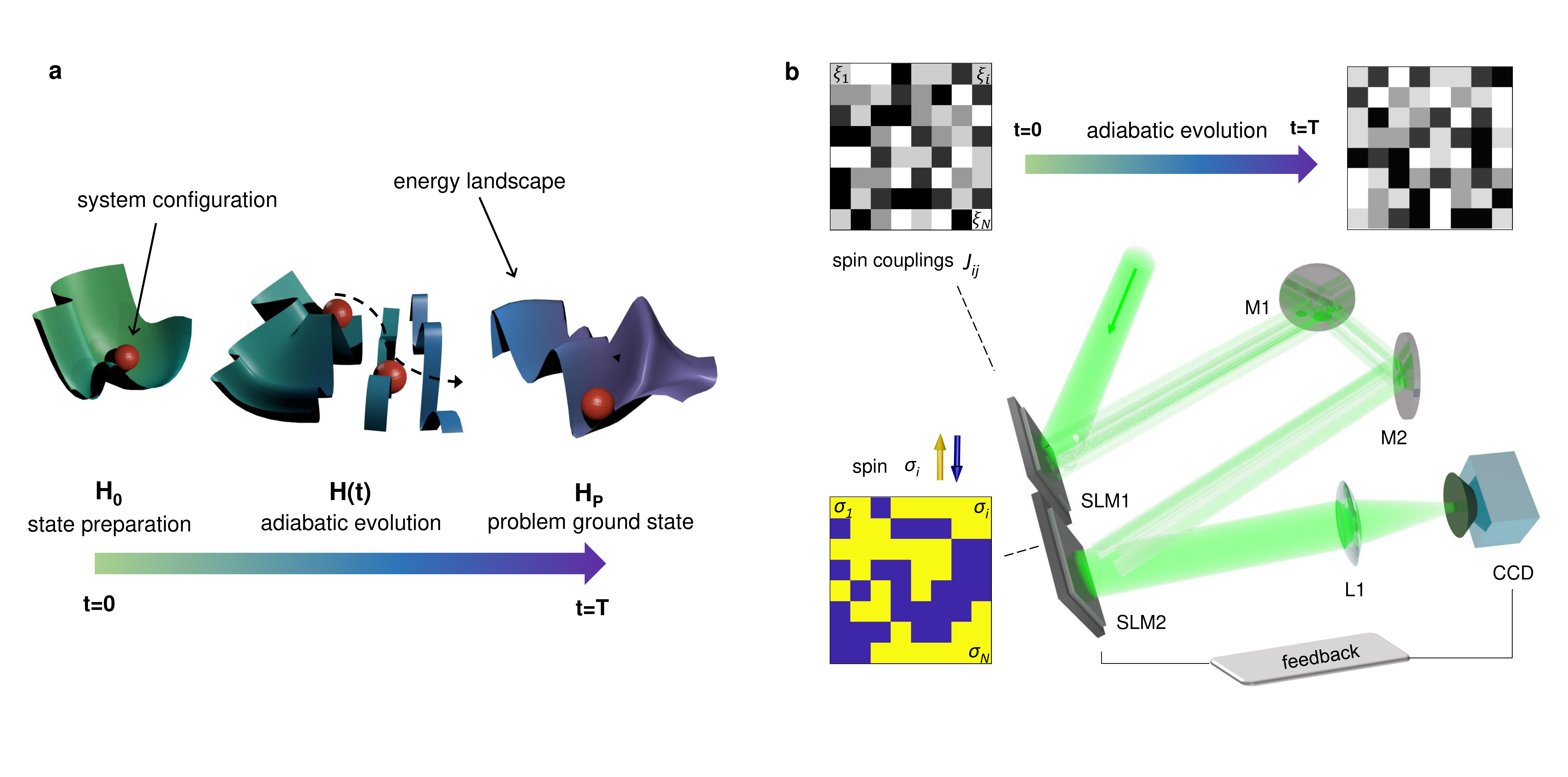} 
\vspace*{-0.6cm}
\caption{{\bf Annealing Ising models on a spatial photonic device.} (a) Illustration of the adiabatic computing principle.
The energetic landscape with several minima varies as the Hamiltonian evolves from the initial $H_0$ to $H_P$ corresponding to the target combinatorial problem. When the dynamics is slow enough (adiabatic condition), the system remains in the low-energy state towards the final ground state, which gives the optimal solution.
(b) Realization of the adiabatic evolution on a spatial photonic Ising machine. In the experimental setup, the spins 
$\sigma_i$ are encoded by SLM2 into binary optical phases (the inset shows a representative configuration), whereas SLM1 modulates the beam
amplitudes $\xi_i$ to control spin interaction (top panels). Adiabatic evolution is implemented by the controlled variation of the couplings via the amplitude-modulated light, as illustrated by the random intensity matrices in the top panels  (M1-M2, mirrors; L1, lens; CCD, camera).
}
\vspace{-0.2cm}
\label{Figure1}
\end{figure*}

In this Article, we demonstrate adiabatic evolution on a spatial photonic Ising machine (SPIM). 
We exploit the features of the SPIM, which encodes the spins and their couplings via spatial light modulators (SLMs).
Starting from the energy minimum of a simple Hamiltonian, and slowly changing the system, we find the low-energy ground state of a target model.
For the adiabatic transformation of  the Ising Hamiltonian, we vary the spin couplings at fixed experimental noise level [Fig.~1(a)]. 
The annealing protocol occurs optically by amplitude modulation. We also test a different holographic annealing scheme, 
which uses the image formed during light propagation to control the instantaneous Hamiltonian.
We consider Mattis spin glasses with $100$-spins initially prepared in a uniform ferromagnetic state. For a sufficiently slow evolution of the Hamiltonian,
the success probability approaches unity. Good performance are maintained also when increasing the number of spins.
Our findings demonstrate a novel approach that allow applying adiabatic computing principles on photonic devices.

\section{Results}

\subsection{Spin dynamics on a spatial-photonic Ising machine}

In a SPIM, a coherent wavefront encodes binary spin variables by spatial light modulation \cite{Pierangeli2019, Pierangeli2020}. The device takes advantage of optical vector-matrix multiplications and of the large pixel density of SLMs, properties that enable implementing large-scale photonic computing and machine learning \cite{Brunner2018, Saade2016, Hamerly2019, Popoff2019, Zuo2019, Marcucci2020, Rafayelyan2020}. Figure~1(b) shows the SPIM with an optical path with two SLMs: SLM1 fixes the couplings of the Hamiltonian by amplitude modulation, and SLM2 controls the spin variables (see Methods). Ising spins $\sigma_{i}=\pm1$ are imprinted on a continuous beam by $0$-$\pi$ phase-delay values (SLM2). The spin interaction occurs by interference on the detection plane.
Spatial modulation of the input intensity (SLM1) fixes the interaction strength. Minimizing the difference between the image detected on the camera and a chosen target image $I_T$ is equivalent to minimizing an Ising Hamiltonian with couplings determined by the amplitude values set by SLM1 and by $I_T$ \cite{Pierangeli2019}.

\begin{figure*}[t!]
\centering
\vspace*{-0.3cm}
\hspace*{-0.2cm} 
\includegraphics[width=1.90\columnwidth]{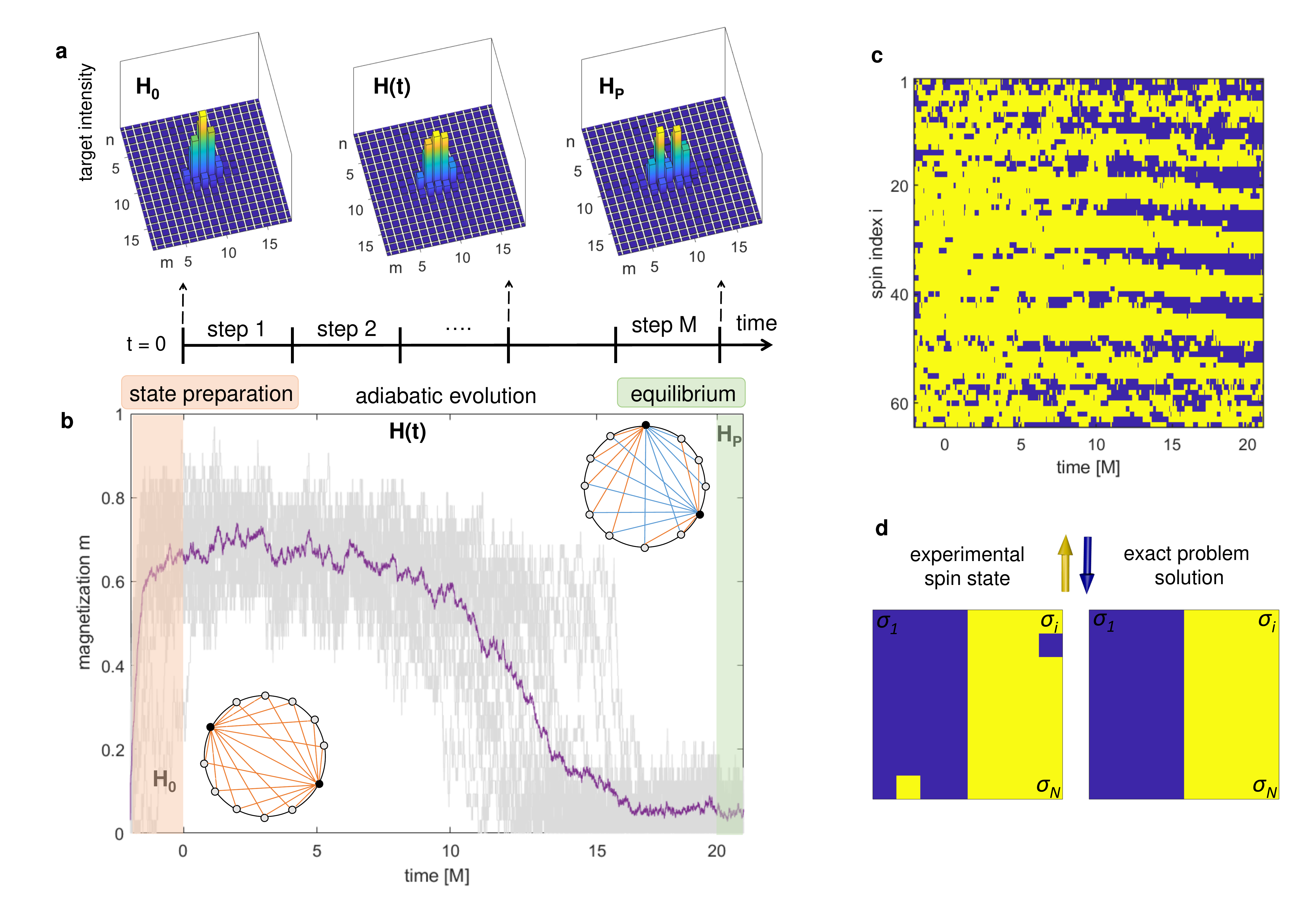} 
\vspace*{-0.2cm}
\caption{ {\bf Optical computing of Mattis spin glasses by holographic annealing.} (a) Evolution of the target image, corresponding to an Ising Hamiltonian  that starts from $H_0$ (state preparation) and evolves towards the problem $H_P$ in $M$ discrete steps. 
(b) Magnetization dynamics for $32$ independent realizations of the annealing process. The thick purple line gives the average evolution ($N=64$, $M=20$). 
Insets show the graphs of $H_0$ and $H_P$. 
(c) Spin configuration as a function of time for a representative Mattis instance. (d) Spin ground state measured at equilibrium and exact combinatorial solution of the realized problem.}
\vspace{-0.1cm}
\label{Figure2}
\end{figure*}

Classical annealing is a well-known strategy to minimize problems having many local minima. A time-dependent Hamiltonian $H(t)$ is made evolving adiabatically from a simple $H(0)=H_0$ to the target problem $H(T)=H_P$ [Fig.~1(a)], 
with $T$ the annealing time. If the evolution is slow enough, the system remains trapped in the ground state of $H(t)$, and a final measurement provides
the spin configuration minimizing $H_P$.
In our implementation, we first perform a state preparation phase in which the system reaches the minimum of a Hamiltonian $H_0$ with homogeneous couplings; then we have the adiabatic evolution, during which the Hamiltonian reaches the target model $H_P$.

{\it State preparation.} The optical machine works in a measurement and feedback loop, with all the parameters kept constant once initialized. Recurrent feedback from the detected intensity allows the phase distribution on the SLM2 to converge towards the ground state of an Ising Hamiltonian $ H_0= - \sum_{ij} J_{ij} \sigma_i \sigma_j$, with couplings $J_{ij} = \xi_i \xi_j \tilde{I}_T(i,j)$  \cite{Pierangeli2019}. The quenched variable $\xi_i$ is the optical amplitude impinging on the $i$-th spin; $\tilde{I}_T$ is the Fourier transform of a pre-determined target image. The difference between $I_T$ and the image detected on the CCD is the cost function. At each measurement and feedback cycle, we update the spin configuration $\{\sigma_i\}$ to minimize the
cost function. Differently from CIMs \cite{Inagaki2016, Vandersande2019}, once initialized, the spin state is updated without electronically computing the energy as well as the field on each spin.

{\it Adiabatic evolution.} To realize annealing, we vary the Hamiltonian with time. During the adiabatic phase, $H(t)$ changes towards the target $H_P$. If the evolution is slow enough to prevent high energy excitations, the final state gives the optimal solution.
We implement the time-dependent Ising Hamiltonian
\begin{equation}
  H (t)= - \sum_{ij} J_{ij}(t) \sigma_i \sigma_j .
  \label{eq:1}
\end{equation}
with the conditions $H(0)=H_0$ and $H(T)=H_P$. The spin couplings $J_{ij}$ can be changed by the target image $I_T$ (``holographic annealing''), and also by varying the modulated amplitudes $\xi_i$ (``optical annealing'' ). This feature enables different and versatile adiabatic evolution protocols. We remark that, due to the experimental noise, the spin system is coupled to an effective thermal bath \cite{Pierangeli2020}.

\begin{figure*}[t!]
\centering
\vspace*{-0.3cm}
\hspace*{-0.2cm}
\includegraphics[width=1.90\columnwidth]{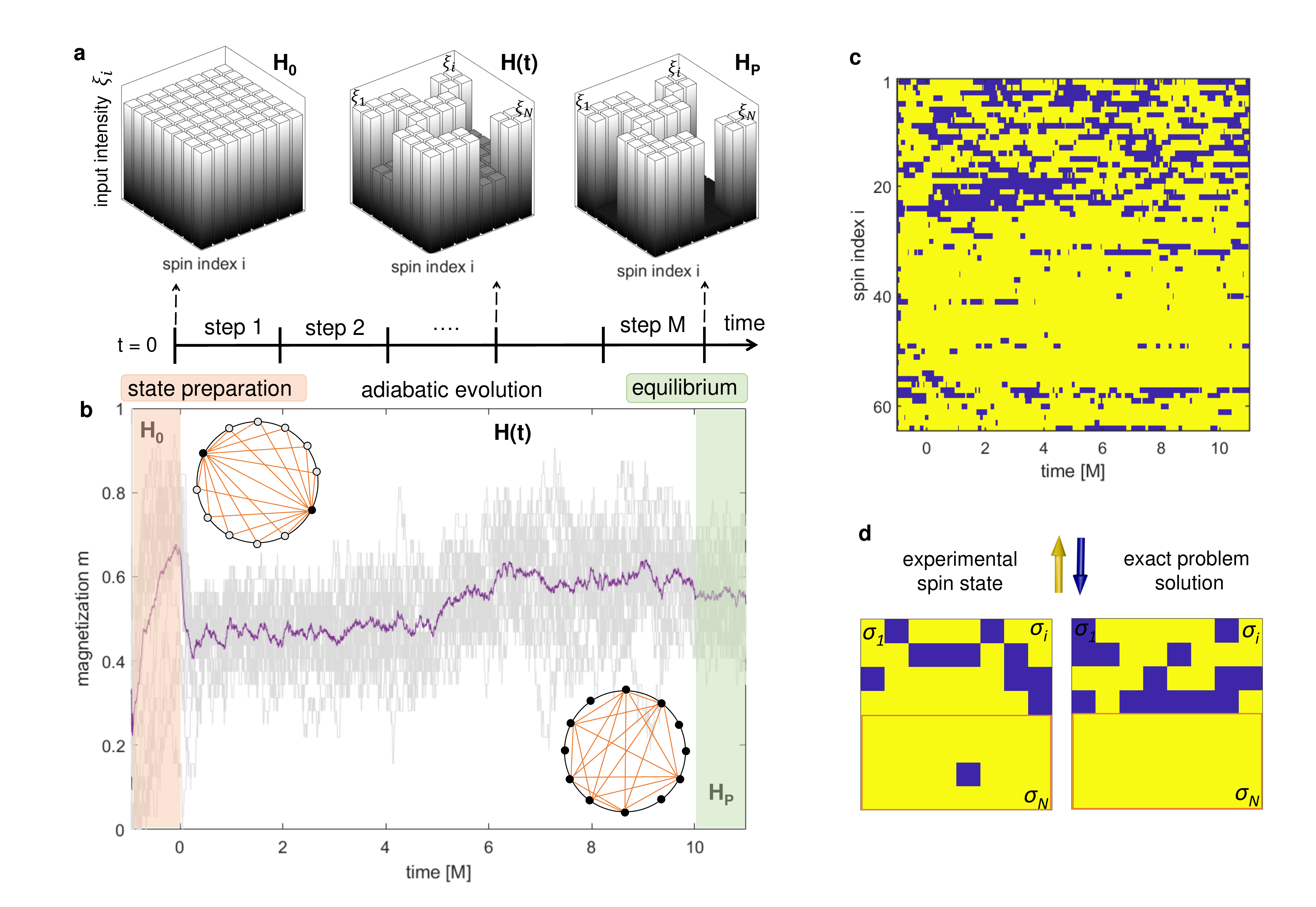} 
\vspace*{-0.2cm}
\caption{{\bf Adiabatic evolution by optical control of the spin couplings.} (a) Variation of the spatially-modulated intensity during the trajectory.
The initial model $H_0$ has uniform couplings;  after $M$ steps, $H(t)$ is equal to the target $H_P$ with random interactions.  (b) Dynamics of the magnetization for replicated experiments with $N=64$ and $M=10$.  The thick purple line gives the average behavior.
The insets show the graphs for $H_0$ and $H_P$. (c) Spin dynamics for a representative realization of couplings. The initial uniform state forms clusters when evolving towards the ground state of the $H_P$. (d) Observed final ground state compared with the exact solution of the corresponding Mattis model.
The orange box indicates the set of interacting spins, while the other part corresponds to those with zero-couplings which fluctuate thermally. 
}
\vspace{-0.1cm}
\label{Figure3}
\end{figure*}

\subsection{Holographic annealing}

A random spin state is prepared in a low-temperature homogeneous ferromagnetic configuration and evolves toward a Mattis spin glass~\cite{Mattis1976, Nishimori2001}. This target $H_P$ belongs to a class of frustrated Ising models whose zero-temperature ground states is characterized by ferromagnetic and anti-ferromagnetic domain blocks. During the evolution, time is discretized in $M$ steps. In each step, the Ising machine performs a fixed number of iterations $n_i$ using the instantaneous Hamiltonian, until the condition $H=H_P$ is reached (see also Methods). The parameter $M$ determines the speed of the adiabatic trajectory, i.e., the total annealing time is $T=n_i M$, following a typical protocol~\cite{Steffen2003}.

In the holographic annealing scheme, we vary the target image $I_T$ to perform the temporal evolution in Eq.~(\ref{eq:1}), as shown in Fig.~2(a).
The graphs for the initial and final problem are inset in Fig.~2(b),
where we report the time evolution of the magnetization $m=\langle\sigma_i\rangle$ for $M=20$. The results correspond to $32$ realizations with $N=64$ spins. The final ground state always has zero mean magnetization, as expected for a spin glass with vanishing mean interaction.
Fluctuations around the average trajectory (thick purple line in Fig.~2(b)) unveil the influence of the experimental noise, which helps the exploration of the various energy configurations during the annealing. 

The considered Mattis models enable to directly test the minimization trajectory, as these specific Ising Hamiltonians admit exact zero-temperature ground states~\cite{Nishimori2001}. Therefore, we can verify the obtained solutions by inspecting the spin configurations. Figure~2(c) shows the dynamics of the spins. Figure~2(d) reports the final ground state configuration and its comparison with the exact solution. Remarkably, the two spin states have the same energy and differ only for two spin flips. This discrepancy is due to the effective thermal fluctuations. The result indicates that the global minimum is successfully found.

\begin{figure*}[t!]
\centering
\vspace*{-0.1cm}
\hspace*{-0.2cm}
\includegraphics[width=1.80\columnwidth]{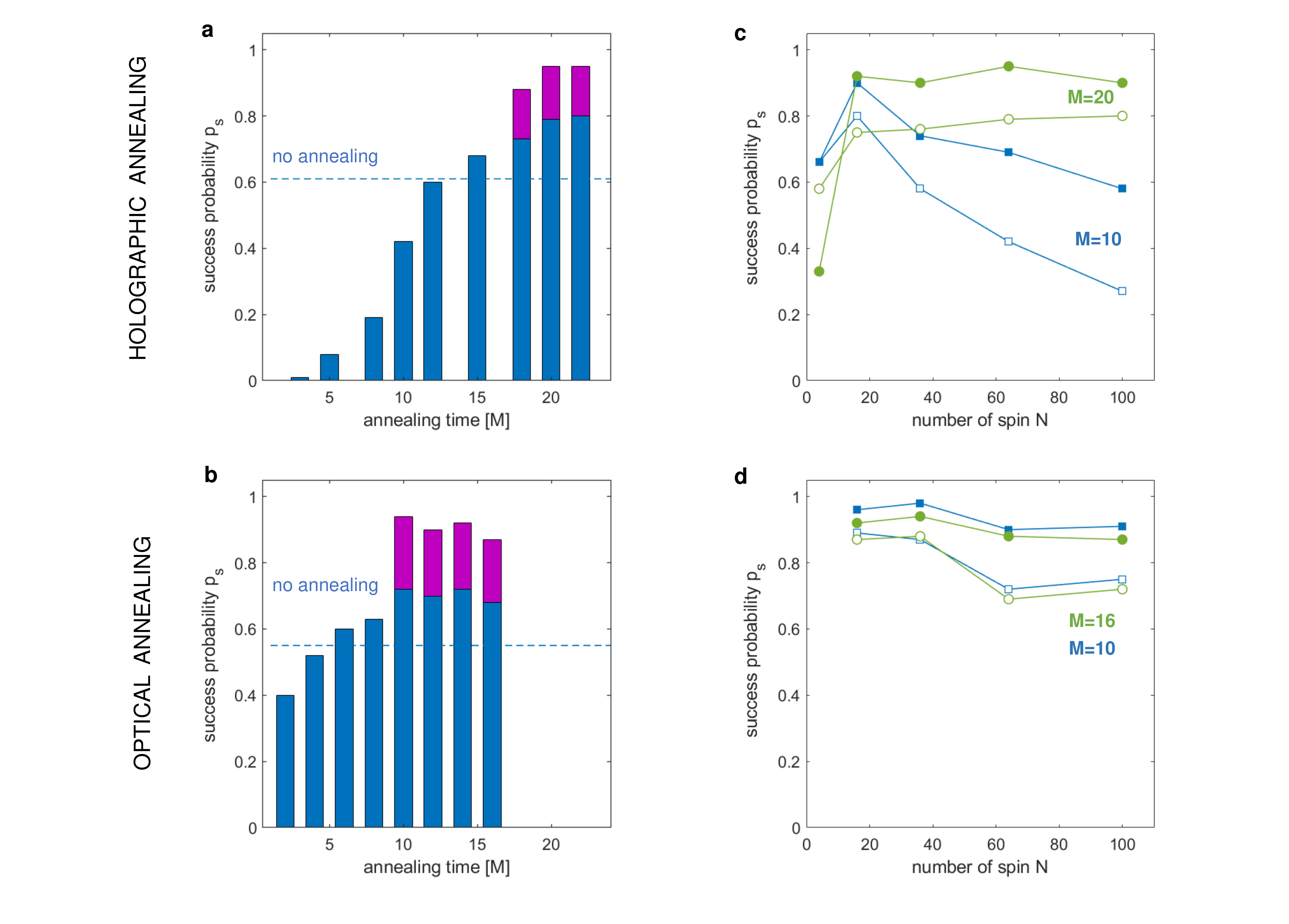} 
\vspace*{-0.3cm}
\caption{ {\bf Optical computing performance and scaling properties.} Success probability as a function of the annealing time for (a) holographic and (b) optical annealing of Ising models with Mattis-type interaction. Blue and magenta bars indicate mean values obtained by measures at a fixed time and averaging over configurations at different times (see Methods). (c, d) Success probability varying the spin number at different $M$. Empty and filled dots are for data from single-shot and noise-averaged measurements.
}
\vspace{-0.1cm}
\label{Figure4}
\end{figure*}

\subsection{Optical annealing}

We implement Eq.~(\ref{eq:1}) by varying the Ising machine parameters optically. Adiabatic evolution is performed by evolving the spatially-modulated intensity inside the photonic machine while keeping constant the target image $I_T$.
In each of the $M$ steps, the machine operates with a different amplitude mask $\xi_i$ (see also Methods). The instantaneous Hamiltonian $H(t)$ is specified by $J_{ij}(t)= \xi_i (t) \xi_j (t)$ (within some multiplicative constant). The initial Hamiltonian $H(t)=H_0$ corresponds to a homogeneous amplitude distribution [Fig.~3(a)]. We choose a random set of amplitudes $\xi$, and gradually decrease their values towards zero. The corresponding system is an instance of a target Mattis model $H_P$ with randomly coupled clusters of spin, as represented by the graph inset in Fig.~3(b). Figure~3(b) shows the time-evolving magnetization when approaching the Mattis ground state during the optical annealing.
Each trajectory is affected by noise-driven fluctuations, and, at the end of the annealing, the expected mean magnetization $m=\pm 0.5$ is reached.
This value arises from fact that half of the spins have $J_{ij}=0$ in $H_P$.
Figure~3(c) shows the evolution of a configuration with $N = 64$ spins for a representative case.
The final ground state averaged over thermal fluctuations coincides with minimum energy state of the programmed Hamiltonian $H_P$.
Figure~3(d) shows the remarkable agreement of the obtained spin configuration with the corresponding theoretical zero-temperature solution.
This demonstrates that the spins maintain their state at the lowest energy during the optical change of the Hamiltonian. 
As reported below, adiabatic evolution of the Ising machine improves the search for the global solution of the encoded optimization problem.

To test the performance of the holographic and optical annealing, we vary the number of discretization steps $M$, i.e., the annealing time. 
The number of machine iterations in each step is kept constant. Figure~4(a) shows the success probability $p_s$ (see Methods) versus $M$ for the holographic annealing; results refer to the problems in Fig.~2 ($N=64$). At a small $M$, when the evolution occurs rapidly, the excitation of high-energy states reduces the effectiveness of the minimization. The probability of converging to the optimal solution increases with the annealing time.
The adiabatic condition is identified by $p_s$ reaching a plateau with values exceeding $90\%$ when averaging over thermal fluctuations (magenta bars in Fig.~4(a), see Methods). Figure~4(b) shows the results for the optical annealing scheme. The adiabatic condition is reached on much shorter annealing time with respect to the holographic case in Fig.~4(a). The ground state is found also for rapid processes ($M\simeq10$).
The quality of each solution is given by its Hamming distance, i.e., the number of spins that should be changed to obtain the known zero-temperature ground state \cite{Boixo2014}. The mean Hamming distance we measure in adiabatic condition is $h=8 \pm 1$ ($h=7.5 \pm 0.8$) for the optical (holographic) annealing. In both methods, we found that adiabatic evolution provides a substantial enhancement of the success probability with respect to a ``no-annealing'' strategy, in which the Hamiltonian  $H(t)=H_P$ is kept constant since the initial instant [dashed lines in Fig.~4(a,b)].
Moreover, Fig.~4(a) and Fig.~4(b) show that - in all the considered cases - the problem ground state can be found more efficiently if the spin fluctuations at equilibrium are observed and averaged out. This circumstance indicates that a large part of the SPIM error can be ascribed to effective thermal noise.

We also investigate the scaling properties of the photonic setting by varying the number of spins. In Fig.~4(c) and Fig.~4(d), we report the scaling of the success probability for holographic and optical annealing, respectively. For a fast evolution protocol [$M=10$, Fig. 4(b)], 
a degrading effect when increasing the size is observed in the holographic case.
However, as the adiabatic condition is reached, we found a remarkable property: good performance are maintained as the problem size grows.
For $N=100$, values of $p_s$ close to unity correspond to a measured mean Hamming fraction ($h/N$) of $0.11$. 
The results suggest that, on these specific problems, adiabatic computing can be extended to larger scales with comparable performance.

\section{Conclusion}

Annealing is one of the most general and consolidated heuristic approaches to solve combinatorial optimization problems and complex physical models.
Its experimental implementation on unconventional physical systems operating at room temperature can impact future computing architectures.
We have realized adiabatic computing schemes on a spatial-photonic Ising machine showing that ground states are found with enhanced success probability.
Computing devices based on spatial light modulation are scalable to larger sizes and can potentially host systems consisting of millions of spins.
They can also benefit from temporal fluctuations to speed-up the search for the optimal minimum \cite{Pierangeli2020}. Recent SLM technologies
can reduce the operation time of our scheme to milliseconds \cite{Tzang2019}.
Other developments may include the use of nonlinear media \cite{Kumar2020}, or metasurfaces, to implement a larger class of combinatorial optimization problems, beyond Mattis instances of the Ising model, and for realizing compact devices without free space propagation.
Exploiting optical matrix multiplications, which can be performed efficiently for large sizes \cite{Rafayelyan2020},
our spatial-photonic Ising machine represents a route to tackle hard optimization problems
at an unprecedented scale, and opens the route to the experimental demonstration of various minimization strategies.

\section{Methods}

\subsection{Experimental setup and feedback method} 
Light from a continuous-wave laser source with wavelength $\lambda= 532$nm (max. power $1.5$W) is expanded, polarization controlled and spatially filtered.
The beam is thus spatially modulated in amplitude by a first spatial light modulator (SLM1) and then it is independently phase modulated by the second modulator (SLM2). The optical path shown in Fig. 1(b) is realized by a single nematic liquid crystal reflective modulator (Holoeye LC-R 720, $1280\times768$ pixels, pixel pitch $20\times20 \mu$m). A section of the modulator is employed in amplitude mode to generate controlled intensity distributions $\xi_i$, which are imaged by a 4-f system [not shown in Fig. 1(b)] and mirrors M1-M2 on the second section, which perform binary phase modulation. By a combination of incident and analyzed polarizations phase-modulation occurs with less than 10\% residual intensity variations. To maintain the setting optically stable, 
an active area of approximatively $200\times200$ SLM pixels is divided into $N$ optical spins by grouping several pixels.
Modulated light is separated using an holographic grating and focused by a lens L1 (f$=500$mm) on a CCD camera.
The intensity is detected on a region of interest composed of $n \times m =18 \times 18$ spatial modes, where the signal in each mode is obtained 
averaging over $10 \times 10$ camera pixels. 
The measured intensity pattern determines the feedback signal. At each iteration a spin is randomly flipped,   
the recorded pattern is compared with a reference image $I_T$ on the same number of modes [see Fig. 2(a)], and the spin configuration on the SLM2
is updated to minimize the difference between the two images. Due to intensity fluctuations, as well as to the grouping procedure at the readout, exists a finite probability to update the spin configuration in any case. We experimentally evaluate this probability to $p\approx0.1$. These are the classical fluctuations through which various energy configurations are explored. The value of noise can be also tuned by applying a post-processing step to the recorded intensity \cite{Pierangeli2020}. However, in all the presented results it is kept fixed at the same level. 

\subsection{Adiabatic evolution schemes}
{\it Holographic annealing.} The machine starts from a random configuration of $N$ spins, and it is prepared on a uniform ferromagnetic state. 
Specifically, the initial Hamiltonian is
 $H_0= -\sum_{ij} \bar{J} \sigma_i \sigma_j$, which is implemented using a plane wave of constant amplitude $\xi_i=E_0$
and a target image that is composed by a single spot [Fig. 2(a)], so that $\tilde{I}_T=c$, being $c$ an arbitrary constant, and $\bar{J}=c E_0^2 $.
By varying in time the target image $I_T$, being $J_{ij} (t) = \xi_i\xi_j \tilde{I}_T(t)$, we implement a linear evolution protocol of the form
$H(t)= (1-t/T)H_0 + (t/T)H_P$, 
with $H_P$ an Ising problem where each spin-spin interaction can have only a positive or a negative value (frustrated Mattis model). 
Therefore, the time-dependent Hamiltonian is
\begin{equation}
H(t)= \left(1-\frac{t}{T}\right)H_0 - c E_0^2 \frac{t}{T}\sum_{ij} G_{ij} \sigma_i \sigma_j,
\end{equation}
where $G_{ij}$ is a block matrix with element values $\pm 1$. For example, a four-block matrix $G=[ G_{11} , G_{12} ; G_{21}, G_{22}]$,
where $G_{11}=G_{22}$ is a all-ones matrix and $G_{12}=G_{21}=-G_{11}$, corresponds to a target image $I_T$ with two horizontal intensity spots [Fig. 2(a)].
In this case, pairs of negatively-coupled spins correspond to points of the optical field resulting in destructive interference on the central mode of the detection plane.
To discretize the evolution, we divide the total time interval $[0,T]$ in $M$ equal intervals. For a fixed system size, we choose to keep constant the number of iterations $n_{i}$ performed by the machine in each step. The annealing time thus increases with $M$.
In Fig. 2 and Fig. 3, we normalize the evolving time $t$ (machine iterations) with respect to $n_{i}$.

{\it Optical annealing.} 
Starting from a random spin state, a uniform ferromagnetic configuration is prepared by implementing $H_0$ as in the holographic protocol. 
However, in the optical evolution method, the target image is maintained fixed to the initial single-spot profile and the whole dynamics is performed by varying the amplitude distribution $\xi_i$ via intensity modulation on the SLM1. This corresponds to individually changing the spin-spin coupling values. 
The evolution protocol we implement is thus
\begin{equation}
H(t)= - c\sum_{ij} \xi_i(t) \xi_j(t) \sigma_i \sigma_j, 
\end{equation}
where $c$ is an arbitrary constant. Since the interaction matrix is given in any case as a product of two variables $J_{ij}\propto \xi_i \xi_j$, 
the problem Hamiltonian $H_P$ still maintains the form of a Mattis model. In this case, each instance corresponds to a random set of positively interacting spins [Fig. 3(b)]. The time dependence of the optical amplitudes $\xi_i(t)$ can also be selected arbitrarily. 
We choose to implement couplings that change linearly in time, that is, to variate the optical intensity with a constant rate.
This linear schedule corresponds to $\xi_i(t) = \sqrt{E_0 - (t/T)E_0}$, and it is shown in Fig. 3(a) for a random set of evolving $\xi_i$.
The total run time $T$ is divided in $M$ equal intervals. Experimentally, the dynamics is implemented by applying an amplitude mask varied in each step. Digital modulation over $256$ grey values ($8$-bit) corresponds to a maximum intensity modulation depth of the order of $10^3$.

\subsection{Ground states analysis}
To quantify the ground state probability at finite temperature, we compute the correlation coefficients between the measured spin configuration
and the known optimal solutions of the programmed instance, which for a Mattis graph is identical to the interaction configuration $\xi_i$, or to its sign reversal \cite{Nishimori2001}. The correlation reads as $C= \sum_{i} \sigma_{i} \xi_i / |\xi_i|$, being $C=\pm 1$ for the ideal spin system in the lowest energy state. It is successful an evolution whose final equilibrium state gives $|C|>0.75$; the success probability $p_s$ is the fraction of runs converging to the correct ground state over a set of experiments with different random initial conditions. Instantaneous values of $p_s$ are obtained from a single measurement at a fixed time (machine iteration) during the equilibrium stage; values averaged over thermal fluctuations (magenta bars in Fig. 4(a,b)) result from averaging on a time interval the spin configuration measured at equilibrium and identifying the spin with the local magnetization sign. The Hamming distance $h$ is the number of spins that need to be flipped to reach the minimum energy configuration. The Hamming fraction is the Hamming distance normalized to the total number of spin, $h/N$. 

\vspace*{0.1cm}
{\bf Acknowledgment.}
We acknowledge funding from SAPIExcellence 2019 (SPIM project), QuantERA ERA-NET Co-fund (Grant No. 731473, project QUOMPLEX), PRIN PELM 2017
and H2020 PhoQus project (Grant No. 820392). We thank Mr. MD Deen Islam for technical support in the laboratory.

\end{document}